\documentclass[runningheads]{llncs}
\usepackage[T1]{fontenc}
\usepackage{graphicx}
\usepackage{amsfonts}
\usepackage{mathtools}

\begin{document}
\title{From Task Allocation to Risk Clearing:\\A Unifying Interface for Mixed Human-Agent Societies}
\titlerunning{Risk-Aware Option Clearing}
%
\author{Vassilis Vassiliades\orcidID{0000-0002-1336-5629}}
\authorrunning{V. Vassiliades}
%
\institute{CYENS - Centre of Excellence, Nicosia, Cyprus\\
\email{v.vassiliades@cyens.org.cy}}

\maketitle              

\begin{abstract}
As humans, robots, and software agents increasingly share safety-critical environments, coordination must move from static task allocation to managing uncertain commitments. Existing frameworks fall short: they either assume rigid, static teams or learn opaque joint policies that are hard to adapt and difficult to integrate with human decision-makers. To overcome these limitations, we propose \emph{Risk-Aware Option Clearing} (ROC), a unifying coordination mechanism in which agents expose options (temporally extended skills) paired with risk summaries that predict outcome distributions. A central clearinghouse then assigns tasks by optimizing risk-adjusted mission utility under deadlines and safety constraints. ROC is a family of mechanisms, ranging from deployments where the clearinghouse learns outcome models from data to ones that consume full distributional predictions from agents. By treating risk-aware options as the basic coordination unit, ROC sketches a scalable, transparent infrastructure for integrating heterogeneous agents into future mixed human--agent societies and outlines a research agenda for such risk-aware clearing layers.

\end{abstract}

\section{Introduction}

When a building is damaged by an earthquake, human medics, inspection drones, and ground robots may all be available to help, but someone still must decide who surveys which stairwell, who clears which corridor, and who carries which kit---and with what confidence that each will finish on time and safely. Across disaster response, energy grids, and city maintenance, the critical question is no longer just ``who does what,'' but \textbf{who does what, when, and with what guarantee of safety and timeliness}. Failing to answer this in a principled way leaves us with brittle, hand-tuned schedules or opaque learned policies that cannot scale to the complexity of mixed human--agent societies.

This challenge creates a coordination problem with three intertwined demands. First, \textbf{openness}: agents are heterogeneous, designed by different entities, and cannot be assumed to share internal representations. Second, \textbf{temporal abstraction}: capabilities are not atomic steps but extended skills that unfold over time. Third, \textbf{explicit risk awareness}: decisions must account for the probability of meeting deadlines and constraints, reasoning about tails rather than just averages. Existing frameworks seldom meet all three simultaneously.

\sloppy Current approaches only partially address this gap. Centralized schedulers and multi-robot task allocation \cite{gerkey2004formal,aziz2021multi} typically assume fixed teams and static capability models, breaking down when teams change or conditions degrade. Multi-agent reinforcement learning (RL) \cite{busoniu2008comprehensive} and recent centralized-training, decentralized-execution methods \cite{rashid2020monotonic,foerster2018counterfactual} offer end-to-end coordination but produce opaque joint policies that are difficult to adapt or audit. Safe and constrained RL \cite{garcia2015comprehensive,wachi2024survey,de2021constrained} tackles constraints in specific formulations, but lacks a general interface for heterogeneous interoperability. Conversely, auction mechanisms \cite{bertsekas1990auction,lagoudakis2005auction,dias2006market,quinton2023market} and the Contract Net Protocol \cite{smith1980contract} support openness but rely on scalar bids that obscure temporal structure \cite{sutton1999between} and risk. Finally, while distributional RL \cite{bellemare2017distributional,rowland2019statistics,bellemare2023distributional} models return uncertainty, it remains focused on single agents rather than on how such predictions should be exposed or combined to coordinate open, heterogeneous teams.

We propose \textbf{Risk-Aware Option Clearing (ROC)} to fill this gap. In ROC, participants (whether humans, robots, or LLM-based assistants) expose \textbf{options} (temporally extended skills) \cite{sutton1999between} paired with \textbf{risk summaries} that predict outcome distributions; together, these form a standardized option--risk interface. A central \textbf{clearinghouse} assigns tasks by trading off mission utility against the risk of missed deadlines or safety violations. By maintaining calibration statistics to penalize overconfidence, ROC functions not as a monolithic algorithm, but as a new unifying \textbf{interface and coordination pattern} for heterogeneous agents. We argue that, just as TCP/IP standardized data exchange, ROC-style protocols could become foundational infrastructure for future mixed human--agent societies, allowing independent agents to plug into a shared, risk-aware decision process.

To situate this research vision, we first formalize the problem of clearing tasks under uncertainty in mixed societies. We then define ROC as a \textbf{family of mechanisms}, ranging from lightweight interfaces where the clearinghouse learns risk models (ROC-Min) to variants that consume full distributional predictions from agents (ROC-Full). Finally, we ground this design in disaster response, micro-grids, and city maintenance, and outline a broader research agenda around standardizing option interfaces, risk summaries, and ROC-style coordinators for future open multi-agent systems.


\begin{figure*}[t!]
    \centering
    \includegraphics[width=1\linewidth]{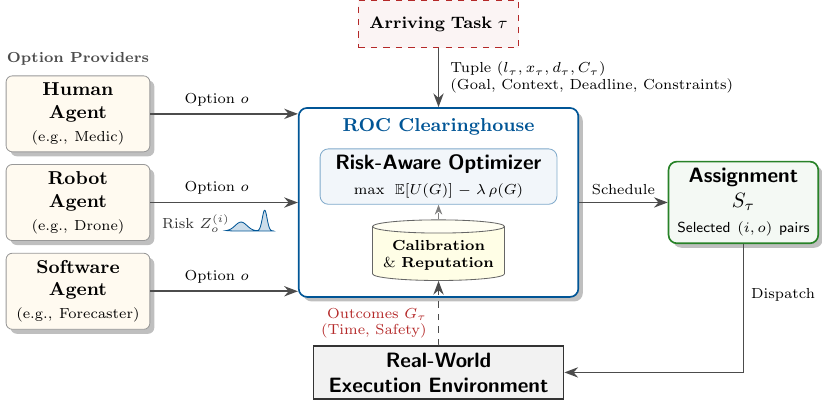}
    \caption{\textbf{The Risk-Aware Option Clearing (ROC) Architecture.} 
    Heterogeneous agents (left) coordinate by exposing capabilities as \emph{Options} ($o$) paired with probabilistic \emph{Risk Summaries} ($Z_o^{(i)}$), rather than scalar bids or costs. 
    The Clearinghouse receives arriving \emph{Tasks} ($\tau$) with explicit constraints and deadlines, and generates a \textbf{Schedule} by maximizing expected utility subject to risk limits (via the optimizer). 
    Selected agents are \textbf{Dispatched} to the execution environment, where realized \emph{Outcomes} ($G_\tau$) provide a feedback loop to update the system's calibration and reputation models.}
    \label{fig:roc_architecture}
\end{figure*}

\section{Risk-Aware Option Clearing (ROC)}
\label{sec:roc}

We model coordination in mixed human--agent societies as online assignment of temporally extended skills to arriving tasks under outcome uncertainty. ROC specifies how agents expose these skills and risk predictions, and how a clearinghouse uses them to decide who does what, when, and with what guarantee.

\subsection{Setting, Tasks, and Outcomes}

We consider a finite set of agents $\mathcal{I} = \{1,\dots,N\}$, which may include humans, robots, and software agents (including modern LLM-based tool-using assistants). At assignment time the system observes a context $x \in \mathcal{X}$, where $\mathcal{X}$ is an abstract space capturing whatever information is available and relevant (location, local environment state, current workload, high-level features, etc.).

Tasks arrive online as tuples $\tau = (\ell_\tau, x_\tau, d_\tau, \mathcal{C}_\tau)$, where $\ell_\tau$ is a goal description (e.g., ``survey stairwell in building A''), $x_\tau \in \mathcal{X}$ is the context at arrival, $d_\tau$ is a deadline or time window, and $\mathcal{C}_\tau$ collects additional constraints such as safety limits, comfort bounds, regulatory requirements, or role/certification conditions. Multiple tasks may be active at once.

\sloppy When some agent $i \in \mathcal{I}$ is assigned to address task $\tau$ using a temporally extended skill, the execution produces a random outcome vector $G_\tau = (T_\tau, S_\tau, R_\tau^{(1)}, R_\tau^{(2)}, \dots)$, where $T_\tau$ is the completion time, $S_\tau \in \{0,1\}$ indicates success with respect to the goal, and $R_\tau^{(k)}$ are task-dependent cost or risk metrics (e.g., peak force, energy usage, comfort violation). ROC is concerned with predicting and trading off these outcomes when making assignments.

\subsection{Options as the Capability Interface}

Each agent $i$ exposes a finite set of temporally extended skills, or \emph{options}, denoted $\mathcal{O}_i = \{o^{(i)}_1,o^{(i)}_2,\dots\}$. Following the options framework~\cite{sutton1999between}, we write an option as $o = (\mathcal{I}_o,\beta_o,\pi_o)$, where $\mathcal{I}_o \subseteq \mathcal{X}$ is an initiation set specifying in which contexts the option can be started, $\beta_o : \mathcal{X} \to [0,1]$ is a termination condition, and $\pi_o$ is an internal policy or controller (for a robot, a low-level control policy; for a human, a protocol; for software, an algorithm). Examples include physical actions such as ``SurveyStairwell'', ``DeliverKit'', or ``ClearCorridor'', informational actions such as ``ComputeForecast'', and human actions such as ``OnSiteTriage'' or ``LicensedInspection''.
We say that option $o$ is \emph{eligible} for task $\tau$ if $x_\tau \in \mathcal{I}_o$ and $o$ respects basic feasibility conditions implied by $\mathcal{C}_\tau$ (for example, that only certified humans may execute a ``LicensedInspection'' option). In ROC, options are the basic units of coordination: rather than scheduling primitive actions, the clearinghouse assigns options to tasks, which provides temporal abstraction and a natural interface for independently designed agents.

\subsection{ROC-Full: Predictions and Decisions}

In the full version of ROC, each agent not only lists its options but also provides risk-aware predictions for their outcomes on a given task. For an eligible pair $(i,o)$ and task $\tau$, agent $i$ supplies an \emph{option-level outcome distribution}
\[
Z^{(i)}_o(x_\tau) \;=\; \mathcal{L}\big(G_\tau \,\big|\, \text{agent } i \text{ executes option } o \text{ for } \tau \text{ from context } x_\tau\big),
\]
that is, a probability distribution over $G_\tau$ under that choice. This can be viewed as a \textit{distributional analogue of a general value function} \cite{sutton2011horde} over mission-relevant quantities such as completion time, success, and risk metrics. In practice, $Z^{(i)}_o(x_\tau)$ is communicated in a finite form, for example a small set of quantiles for $T_\tau$ and relevant $R_\tau^{(k)}$ together with a success probability $\mathbb{P}[S_\tau = 1]$, optionally augmented with a cost estimate $c_i(\tau,o)$ (e.g., expected energy or internal resource usage).

For task $\tau$, let $\mathcal{A}_\tau = \{(i,o) : i \in \mathcal{I},\, o \in \mathcal{O}_i \text{ eligible for } \tau\}$ denote the set of candidate agent--option assignments. The clearinghouse selects a subset $\mathcal{S}_\tau \subseteq \mathcal{A}_\tau$ (e.g., a primary option and a backup). The induced outcome depends on this composition: we denote by $G_\tau(\mathcal{S}_\tau)$ the random outcome vector under a policy that specifies how options in $\mathcal{S}_\tau$ are used (e.g., with a backup executing only if the primary fails or times out). We write $U\big(G_\tau(\mathcal{S}_\tau)\big)$ for the mission utility, which is large when the task succeeds, completes before its deadline $d_\tau$, and remains within the relevant safety or comfort requirements specified in $\mathcal{C}_\tau$. We write $\rho\big(G_\tau(\mathcal{S}_\tau)\big)$ for a risk measure applied to this random outcome, such as the probability of deadline violation or a Conditional Value-at-Risk (CVaR) \cite{rockafellar2000optimization} over selected safety metrics $R_\tau^{(k)}$.

For the current set of active tasks $\mathcal{T}$, we write $\mathbf{S} = \{\mathcal{S}_\tau\}_{\tau \in \mathcal{T}}$ for a global assignment. ROC-Full defines the clearinghouse decision as
\[
\mathbf{S}^* \;\in\; \arg\max_{\mathbf{S}} \sum_{\tau \in \mathcal{T}} \Big( \mathbb{E}\big[U\big(G_\tau(\mathcal{S}_\tau)\big)\big] \;-\; \lambda\,\rho\big(G_\tau(\mathcal{S}_\tau)\big) \Big)
\]
subject to the per-task constraints $\mathcal{C}_\tau$ (e.g., that $\mathbb{P}[T_\tau \le d_\tau]$ or $\mathbb{P}[R_\tau^{(k)} \le r_{\max}^{(k)}]$ exceed required thresholds) and global resource constraints (e.g., shared battery capacity or operator attention limits across tasks). The parameter $\lambda \ge 0$ encapsulates the risk appetite of the system: smaller values prioritize expected mission utility, while larger values enforce stronger conservatism (for instance, a higher $\lambda$ during a ``Code Red'' disaster state).
In the simplest case where exactly one $(i,o)$ must be chosen per task, this reduces to computing a risk-adjusted score for each candidate in $\mathcal{A}_\tau$ and picking the best. With multiple simultaneous tasks and shared resources, the problem becomes a stochastic assignment and robust scheduling problem over subsets of $\mathcal{A}_\tau$ and tasks in $\mathcal{T}$, typically solved approximately \cite{burkard2009assignment,powell2011approximate,herroelen2004robust,pinedo2022scheduling}. ROC specifies this decision structure but remains agnostic to the particular combinatorial solver used.

The clearinghouse evaluates and compares the predicted outcomes for different candidate selections $\mathcal{S}_\tau$ using the option-level distributions $Z^{(i)}_o(x_\tau)$ supplied by the agents. Over time it maintains reputation and calibration statistics for each agent and option by comparing reported predictions to realized outcomes $G_\tau$. These statistics can be used to bias future decisions (e.g., preferring better calibrated agents or requiring larger safety margins from overconfident ones) or to internally adjust the effective outcome models used in the objective above.


\subsection{ROC-Lite/Min: Simplified Interfaces}

In many deployments it may not be feasible or necessary for agents to provide full outcome distributions. ROC is therefore best understood as a family of mechanisms that share the same structure but differ in how much predictive information agents expose and how much is learned centrally.

In \emph{ROC-Lite}, agents provide simplified risk summaries for each eligible option instead of full $Z^{(i)}_o(x_\tau)$. An agent may report an expected completion time $\hat{m}_o^{(i)}(x_\tau) \approx \mathbb{E}[T_\tau]$, a success probability $\hat{p}_o^{(i)}(x_\tau) \approx \mathbb{P}[S_\tau = 1]$, and a few quantiles or coarse risk classes for deadlines and key metrics. The clearinghouse approximates the ROC-Full objective from these summaries, while the assignment problem over $\mathcal{A}_\tau$ remains the same.

In \emph{ROC-Min}, agents do not provide explicit risk predictions. They simply advertise their capabilities by exposing the option sets $\mathcal{O}_i$ and basic metadata such as roles, locations, and nominal costs. The clearinghouse learns empirical outcome models $\tilde{Z}^{(i)}_o(x) \approx \mathcal{L}(G_\tau \mid i,o,x_\tau = x)$ from logged executions and uses these models in place of $Z^{(i)}_o$ when solving the same assignment problem. ROC-Min is thus the easiest tier to deploy: agents need only agree to the option interface, while risk modelling is handled centrally.

Across ROC-Full, ROC-Lite, and ROC-Min, the underlying pattern is identical: agents expose options as standardized, temporally extended capabilities; for each task the clearinghouse gathers some form of risk information about eligible options (from agents, from its own models, or both); and it selects assignments by explicitly trading off expected mission utility against risk under deadlines and constraints. Figure~\ref{fig:roc_architecture} visualizes this complete architecture: heterogeneous agents (human, robot, software) transmit options and risk summaries to the central clearinghouse, which optimizes the schedule and dispatches assignments, closing the loop by feeding realized outcomes back into the calibration database.


\section{Illustrative Scenarios}
\label{sec:scenarios}

We next illustrate how ROC can be instantiated in three qualitatively different domains. The goal is not to fully specify deployments, but to show how the same coordination pattern---options plus a risk-aware clearinghouse---adapts to fast, safety-critical response, slower continuous control, and privacy-sensitive city-scale management.

\subsection{Disaster Response}

Consider a structural failure in a dense urban area where human medics, inspection drones, and ground robots must coordinate under tight constraints. Tasks correspond to concrete goals such as ``survey stairwell S within six minutes.'' Agents expose options reflecting their capabilities (e.g., a drone offering \texttt{SurveyStairwell}, a medic offering \texttt{OnSiteTriage}), and an \textbf{edge ROC instance}---running in a command vehicle or local control center---acts as the clearinghouse. In a \textit{ROC-Lite or ROC-Full} configuration, it broadcasts tasks; eligible agents respond with risk summaries or outcome distributions $Z^{(i)}_o(x_\tau)$ predicting completion time, success probability, and safety risks.
\textit{Example:}
For $\tau=(\textit{survey stairwell S},x_\tau,6\text{ min},\mathcal{C}_\tau)$, where $\mathcal{C}_\tau$ requires a usable map without entering unsafe zones, a drone exposes $o=\texttt{SurveyStairwell}$ with metadata such as $\mathcal{I}_o=\{\text{within 80m, battery}>30\%,\text{link available}\}$ and termination on map return or timeout. In \textit{ROC-Lite}, it may report $\Pr(S_\tau=1)=0.90$, $\Pr(T_\tau\leq 6)=0.82$, and a smoke-risk flag; in \textit{ROC-Full}, these are marginals of $Z^{(i)}_o(x_\tau)$ over $(T_\tau,S_\tau,R_\tau^{\text{safety}},\ldots)$. If execution yields $G_\tau=(5.4\text{ min},1,\text{safe})$, the clearinghouse logs this outcome with prior executions of $(i,o)$ and updates calibration/reputation statistics.
The clearinghouse might assign a fast but riskier drone as the primary option and a slower but safer ground robot as backup, choosing the pair $\mathcal{S}_\tau$ that maximizes the risk-adjusted objective in Section~\ref{sec:roc}. For small $\lambda$ it favors plans that complete quickly on average, while for larger $\lambda$ it prefers plans whose predicted completion times stay within the deadline $d_\tau$ with high probability. As tasks repeat, the clearinghouse refines empirical outcome models. New robot fleets or human teams can integrate immediately: once they expose options and risk summaries, they can participate without the central controller understanding their internal mechanics.

\subsection{Micro-Grid Energy Management}

In distribution-level energy systems, operators coordinate buildings, batteries, and EV fleets over horizons of minutes to hours. Tasks range from ``limit the feeder peak'' to ``provide frequency support,'' with constraints $\mathcal{C}_\tau$ capturing comfort bounds (e.g., building temperature) and device limits. Agents expose options (e.g., a building offering \texttt{PreCool} or \texttt{ShedLoad}, or a battery offering \texttt{Discharge}) paired with risk summaries. In a \textit{ROC-Lite or ROC-Full} configuration, each agent reports outcome statistics for eligible options, such as predicted power profiles and the probability of violating state-of-charge or comfort constraints, instead of simple scalar costs.
A \textbf{cloud-based ROC} clearinghouse (e.g., at the distribution system operator) aggregates these summaries to select a portfolio of options that jointly meets grid objectives. The ROC objective balances expected peak reduction or cost savings against the risk of comfort or device violations. While slower timescales allow more elaborate optimization than in disaster response, the coordination pattern remains the same: independent agents present options plus probabilistic predictions, and the clearinghouse optimizes for collective success within acceptable risk margins.

\subsection{City Maintenance}

City-scale maintenance introduces longer time horizons and strict privacy boundaries. Tasks range from ``inspect elevator in building X'' to ``clean graffiti at location Y,'' often requiring coordination between private building management systems and municipal contractors. We propose a \textbf{hierarchical deployment} to handle this complexity. At the building level, a \textit{ROC-Min} instance coordinates local robots and staff using internal data. Building agents expose options (e.g., \texttt{InspectElevator}) without sharing raw sensor streams; instead, the local clearinghouse learns empirical outcome models $\tilde{Z}^{(i)}_o(x)$ from logs and aggregates them into coarse risk summaries (e.g., ``completable within 3 days with high probability'').
At the city level, a \textit{ROC-Lite} instance receives these summaries alongside option reports from municipal crews and assigns high-level tasks by trading off utility against risk, while respecting the autonomy of lower-level entities.
This hierarchical combination demonstrates ROC's flexibility: local entities retain control over policies and privacy, yet still participate in city-wide, risk-aware coordination by exposing standardized option capabilities and approximate risk information through a common interface.
Thus different ROC tiers (Min locally, Lite centrally) can coexist in one deployment.

Across these domains, ROC acts as a common interface: different clearinghouse algorithms coordinate heterogeneous agents through risk-aware options.


\section{Discussion and Outlook}
\label{sec:discussion}

ROC bridges established subfields of multi-agent systems (MAS). 
It extends task allocation~\cite{gerkey2004formal} with risk-aware option reports, and is related to the Contract Net Protocol~\cite{smith1980contract}: rather than selecting among scalar bids or proposals, it selects among standardized option--risk reports.
It complements Multi-Agent RL~\cite{busoniu2008comprehensive} by decoupling training from coordination: rather than solving a decentralized decision problem over primitive actions, ROC centralizes coordination while leaving option execution and prediction local to independently designed agents. It also lifts concepts from hierarchical and distributional RL~\cite{sutton1999between,bellemare2017distributional} from internal agent machinery to inter-agent messages.
Finally, it aligns with normative MAS~\cite{boella2006introduction,dignum2001modelling}: roles, certifications, and safety policies can be encoded in $\mathcal{C}_\tau$ and $U$, while the clearinghouse provides the quantitative backbone for risk-aware assignment.

The ROC tiers expose a practical engineering trade-off. \textit{ROC-Min} is easiest to deploy, because agents only expose capabilities, but it is cold-start limited and depends on the clearinghouse learning outcome models from logs. \textit{ROC-Lite} adds lightweight risk summaries, making deadline and safety reasoning possible with modest communication and modelling burden. \textit{ROC-Full} is the most expressive, supporting tail-risk reasoning and richer portfolios, but requires calibrated predictive agents and more complex optimization. In all tiers, dynamic coordination is handled by repeatedly re-clearing the active task set $\mathcal{T}$ as new tasks, outcomes, or failures arrive; parallel execution, task dependencies, shared resources, and operator attention limits are expressed as constraints in the global assignment and dispatch problem, while backups can be activated when predicted or observed risk increases. These benefits come with limitations: ROC depends on a shared option-and-risk interface, sufficient communication for timely re-clearing, calibrated outcome models, and solvers fast enough for the deployment timescale.

A natural evaluation path is to instantiate the disaster-response scenario as a stochastic task-allocation benchmark with heterogeneous drones, robots, and human proxies. ROC-Min/Lite/Full could be compared against scalar-cost auctions, Contract Net variants, and centralized schedulers under varying task arrival rates, agent failures, prediction quality, and risk appetite $\lambda$. Relevant metrics include mission success, deadline-violation rate, safety-violation rate, risk-adjusted utility, reassignment latency, communication overhead, robustness to adding/removing agents, and
calibration error between reported predictions and realized outcomes, using proper scoring rules such as the Brier score for event probabilities and CRPS for full predictive distributions~\cite{gneiting2007strictly}.
Such experiments would test not only whether ROC improves allocation quality, but also when richer risk summaries justify their additional engineering cost.

Several concrete research directions follow. First, a minimal ``ROC API'' must specify which option metadata, outcome signals, quantiles, confidence levels, and temporal dependencies should be standardized. Second, learning architectures must determine how prediction is split between agents and the clearinghouse, especially under concept drift, degrading hardware, or unfamiliar contexts. Third, human participants require interfaces that expose their actions as options without cognitive overload; a digital proxy, such as a tablet or wearable, could translate actions like triage or inspection into option reports while allowing humans to query or override assignments. Fourth, safety-critical deployments should combine probabilistic clearing with formal safety mechanisms such as runtime shielding~\cite{alshiekh2018safe}. Finally, in competitive settings agents may under-report tail risks; strictly proper scoring rules~\cite{gneiting2007strictly} and prediction-market mechanisms~\cite{wolfers2004prediction,chakravorti2023artificial} offer one route to incentivizing calibrated risk reports.

Taken together, these directions position ROC not as a fixed algorithm but as an engineering pattern for mixed human--agent societies. By treating temporally extended skills and risk-aware predictions as first-class interface objects, ROC offers a reusable clearing layer through which heterogeneous agents can coordinate under deadlines, constraints, and uncertainty.



%
%
%

\bibliographystyle{splncs04}

\end{document}